\documentclass[5p]{elsarticle}

\usepackage{graphicx}
\usepackage{dcolumn}
\usepackage{bm}
\usepackage{hyperref}
\usepackage[T1]{fontenc}
\usepackage[utf8]{inputenc}
\usepackage{pslatex}
\usepackage{textcomp} 
\usepackage{color}
\usepackage{lmodern} 
\usepackage[normalem]{ulem}
\usepackage{amsmath}
\usepackage{amsfonts}
\usepackage{breakurl} 
\usepackage{array,multirow,graphicx}
\usepackage{lscape}

\hypersetup{
    breaklinks = true,
    colorlinks = true,
    pdftitle = {}, 
    pdfauthor = {},
    pdfkeywords = {},
    linkcolor = blue,
    citecolor = blue,
    filecolor = black,
    urlcolor = magenta
}

\makeatletter 
\def\@author#1{\g@addto@macro\elsauthors{\normalsize%
    \def\baselinestretch{1}%
    \upshape\authorsep#1\unskip\textsuperscript{%
      \ifx\@fnmark\@empty\else\unskip\sep\@fnmark\let\sep=,\fi
      \ifx\@corref\@empty\else\unskip\sep\@corref\let\sep=,\fi
      }%
    \def\authorsep{\unskip,\space}%
    \global\let\@fnmark\@empty
    \global\let\@corref\@empty  
    \global\let\sep\@empty}%
    \@eadauthor={#1}
}
\makeatother

\graphicspath{{./figures/}}

\newcommand{\lani}{La$_2$Ni$_7$} 
\newcommand{\lamgni}{La$_{2-x}$Mg$_x$Ni$_7$} 
\newcommand{\lamghalf}{La$_{1.5}$Mg$_{0.5}$Ni$_7$}

\begin{document}
\begin{sloppypar}

\title{Effect of substitution La by Mg on electrochemical and electronic properties in \lamgni{} alloys: a combined experimental and \textit{ab initio} studies}

\author{Miros\l{}aw Werwi\'nski\corref{cor1}}
\cortext[cor1]{Corresponding author} 
\ead{werwinski@ifmpan.poznan.pl}

\author{Andrzej Szajek}
\author{Agnieszka Marczyńska}
\author{Lesław Smardz}

\address{Institute of Molecular Physics, Polish Academy of Sciences,\\  M. Smoluchowskiego 17, 60-179 Pozna\'n, Poland}

\author{Marek Nowak}
\author{Mieczysław Jurczyk}

\address{Institute of Materials Science and Engineering, Poznań University of Technology,\\
Jana Paw\l{}a II No. 24, 61-138 Poznań, Poland}

\begin{abstract}
La-Mg-Ni-based alloys are promising negative electrode materials for 3rd generation of Ni-MH$_x$ batteries.
In this work we investigate the effect of Mg substitution on the electrochemical and electronic properties of \lamgni{} materials. 
The mechanical alloying technique is used to produce a series of \lamgni{} alloys ($x$~=~0.00, 0.25, 0.50 and 0.75). 
The X-ray diffraction measurements indicates multi-phase character of the samples with majority (La,Mg)$_2$Ni$_7$ phases of hexagonal Ce$_2$Ni$_7$-type and rhombohedral Gd$_2$Co$_7$-type.
Electrochemical measurements show how the maximum discharge capacity ($C_{max}$) increases with Mg concentration and 
that reach the highest value of 304~mAh/g for \lamghalf{} ($x$~=~0.5). 
The experimental efforts are followed by the density functional theory (DFT) calculations performed with the full-potential local-orbital minimum-basis scheme (FPLO). 
To simulate chemical disorder we use the coherent potential approximation (CPA). 
The calculations are focused on the \lamghalf{} composition with the highest measured value of $C_{max}$.
Additionally, several other structures is considered as reference points.
We find that hexagonal and rhombohedral structures of \lani{} have almost identical total energies which is in a good agreement with a coexistence of both phases in the samples.
The calculated site preferences of Mg in both Ce$_2$Ni$_7$-type and Gd$_2$Co$_7$-type \lamghalf{} phases are consistent with the previous experimental data.
Furthermore, the valence band of the nanocrystalline \lamghalf{} sample is investigated by X-ray photoelectron spectroscopy (XPS). 
The experimental XPS are interpreted based on the corresponding spectra calculated with DFT.
\end{abstract}

\date{\today}

\maketitle

\section{Introduction}\label{sec:introduction}

%
Metal hydrides (MH$_x$) are one of the most important alternatives for hydrogen storage applications and
have a big potential to solve several energy and environmental issues~\cite{varin_nanomaterials_2009}.
Metal hydrides can reversibly absorb and desorb hydrogen at ambient pressure and temperature.
The microcrystalline hydride materials are conventionally prepared by arc or induction melting and subsequent annealing. 
The hydriding-dehydriding properties of metal hydrides can be improved by introduction of metastable phases and formation of nanocrystalline structures.
It can be achieved through the application of a non-equilibrium processing technique, like for example mechanical alloying (MA)~\cite{li_characterization_2014, nowak_nanotechnology_2017}.
From a perspective of industrial application, an important advantage of the MA process is the fact that
the MA allows for production of large quantities of the material, keeping relatively low costs.

%
The recently investigated metal hydrides are the ternary microcrystalline compounds (RE-Mg)$_2$Ni$_7$ (RE = rare-earth metals)~\cite{liu_phase_2016, zhang_electrochemical_2015, crivello_first_2015, liu_enhanced_2015, zhang_hydrogen_2016}.
Their hydrogen storage properties proved to be better than the characteristics of the corresponding binary compounds AB$_n$ ($2 \leq n \leq 5$)~\cite{liu_phase_2016}.
(RE-Mg)$_2$Ni$_7$ alloys with La or Ce are characterized by high hydrogen capacity, moderate hydrogen equilibrium pressure and relatively inexpensive constituent elements.
An open question remains the influence of Mg substitutions in place of RE on the structural, electrochemical and electronic characteristics of (RE-Mg)$_2$Ni$_7$ alloys.

%
This work continues our previous research on La-Mg-Ni alloys and composites~\cite{jurczyk_mg-based_2011, jurczyk_hydrogen_2012, balcerzak_hydrogenation_2017}.
An objective of this study is to investigate the influence of Mg concentration on the electrochemical and electronic properties of \lamgni{} materials.
The series of \lamgni{} alloys (with $x$~=~0.00, 0.25, 0.50 and 0.75) is produced by the MA technique.
The experimental efforts of materials preparation and characterization are followed by theoretical studies based on the density functional theory (DFT).
%
%
The (La,Mg)$_2$Ni$_7$ systems have been investigated by DFT before.
Crivello \textit{et al.}~\cite{crivello_structural_2011} presented a systematic study of every ordered configuration of (La,Mg)$_2$Ni$_7$ and several other La-Mg-Ni systems.
Based on the calculated heats of formation Crivello \textit{et al.} concluded that the stability of the (La,Mg)$_2$Ni$_7$ ternary system decrease with Mg substitution,
wherein both rhombohedral (Gd$_2$Co$_7$-type) and hexagonal (Ce$_2$Ni$_7$-type) symmetries were considered.
In the next paper Crivello \textit{et al.}~\cite{crivello_first_2015} present a systematic DFT study on distribution of hydrogen in La$_2$Ni$_7$, Mg$_2$Ni$_7$ and \lamghalf{} hosts,
however the calculations were limited to the Ce$_2$Ni$_7$ prototype.
In both mentioned works the Mg alloying was modeled based on the ordered compound method.
Furthermore, as the surveys focus on the systems stability, no results of valence band investigations were presented.
The valence band of isostructural hexagonal La$_2$Co$_7$ has been shown before.~\cite{kuzmin_magnetic_2015}

\section{Experimental and Computational Details}
%
%
This sections covers the details of preparation of \lamgni{} samples and reference elemental thin films.
The applied characterization techniques, which are X-ray diffraction (XRD), X-ray photoelectron spectroscopy (XPS) and electrochemical measurements are also described.
Finally, the computational parameters of \textit{ab initio} calculations are given.
\subsection{Materials Preparation}
For preparation of the samples we used the 
La powder -- grated from rod (Alfa Aesar, 99.9\%), Mg powder (Alfa Aesar, -325 mesh, 99.8\%), and Ni powder (Aldrich, 5~$\mu$m, 99.99\%) were used.
\lamgni{} ($x$~=~0.00, 0.25, 0.50, and 0.75) alloys powders were prepared by the MA and annealing process. 
The MA was performed using the SPEX 8000 Mixer Mill. 
A protective argon atmosphere was applied. 
Elemental powders (La, Mg, Ni) were weighted, blended and poured into vials in glove box (Labmaster 130) filled with controlled argon atmosphere (O$_2$~<~2~ppm and H$_2$O~<~1~ppm). 
A composition of starting materials mixture was based on a stoichiometry of an \textit{ideal} reaction. However, due to oxidation of La and Mg the content of theses element was increased by 8 wt\%. 
The amount of La and Mg extra addition (8 wt\%) was determined during our basic research (not shown here), 
in order to obtain after MA process and annealing, materials with chemical composition as close as it is possible to stoichiometry of an \textit{ideal} reaction.
In all cases the MA process lasted for 48~h in argon atmosphere. 
In all cases the as-milled materials were heat treated in high purity argon atmosphere at 1123~K for 0.5~h.
Furthermore for the reference reasons in XPS characterization 
the standard La, Mg, and Ni thin films with thicknesses of about 200~nm were prepared at room temperature using computer-controlled UHV magnetron co-sputtering~\cite{smardz_structure_2000}.
Ni and Mg (La) targets were sputtered using DC and RF modes, respectively. 
The base pressure before the deposition process was lower than 5~$\times$~10$^{-10}$~mbar. 
As a substrate we used Si(100) wafers with an oxidized surface to prevent a silicide formation. 
Therefore we applied a special heat treatment in ultra high vacuum (UHV) before deposition in order to obtain an epitaxial SiO$_2$ surface layer~\cite{deal_physics_2013, skoryna_modification_2015}.
The distance between sputtering targets and substrate was about 220~mm. 
Typical sputtering conditions are listed in Tab.~\ref{tab:sputtering}.
The chemical composition and the cleanness of all layers was checked \textit{in situ} immediately after deposition.
The samples were transferred to an UHV (4~$\times$~10$^{-11}$~mbar) analysis chamber equipped with XPS, Auger electron spectroscopy (AES) and ion gun etching system.
\begin{table}[!ht]
\caption{\label{tab:sputtering} Typical sputtering conditions used for the deposition of La, Mg and Ni thin films.\\}
\centering
\begin{tabular}{|p{3cm}|c|p{1.5cm} | p{1.5cm}|}
\hline \hline
Parameter 				& Unit 	&    La, Mg       				&    Ni       	\\
\hline
Rest gas pressure 			& mbar 	& \multicolumn{2}{c|}{$5 \times 10^{-10}$} 			\\
\hline
Argon partial pressure 			& mbar 	& \multicolumn{2}{c|}{$1 \times 10^{-3}$} 			\\
\hline
Argon purity 				& \% 	& \multicolumn{2}{c|}{99.9998} 					\\
\hline
Target diameter 			& mm 	& \multicolumn{2}{c|}{51} 					\\
\hline
Target purity 				& \% 	& 99.95          				& 99.95        	\\
\hline
Distance between substrate and target 	& mm 	& \multicolumn{2}{c|}{220} 					\\
\hline
Sputtering method 			& - 	& Magnetron RF          			& Magnetron DC	\\
\hline
Sputtering power 			& W 	&  40-60         				& 30-50        	\\
\hline
Deposition rate 			& nm/s 	&  0.01-0.07         				& 0.01-0.1    	\\
\hline
Substrate temperature during deposition & K 	& \multicolumn{2}{c|}{295}					\\
\hline \hline
\end{tabular}              
\end{table} 
\subsection{Structural Characterization}
The crystallographic structures of the \lamgni{} samples were investigated at room temperature using the Panalytical Empyrean XRD with Cu\,K$_{\alpha_1}$ ($\lambda = 1.54056$~\AA{}) radiation. 
The conditions of XRD measurements were: voltage 45~kV, anode current 40~mA and 2~Theta range 20\textdegree--80\textdegree.
\subsection{Electrochemical Measurements}
Mechanically alloyed and annealed materials, in nanocrystalline form, with 10~wt.\% addition of Ni powder, were used to form metal-hydride electrodes. 
Materials were pressed (1050~MPa) with nickel nets acting as current collector to 0.5~g pellets. 
The diameter of each electrode was 8.0~mm and the weight was approximately 0.3~g. 
Initial activation was carried out by soaking of the electrode in 6~M~KOH for 24~h at room temperature.
Electrochemical measurements were made using the Multi-channel Battery Interface ATLAS 0461 and ATLAS 0961. 
Studies were done in an open three-compartment glass cell, using a much larger NiOOH/Ni(OH)$_2$ counter electrode and a mercury oxide (Hg/HgO/6~M~KOH) reference electrode.
The electrodes were charged and discharged at 40~mA/g and the cut-off potential $vs$ Hg/HgO/6~M~KOH amounts to -0.7~V. 
All electrochemical measurements were carried out in deaerated 6~M~KOH solution prepared from pure KOH and 18~M$\Omega$/cm water, at temperature of about 293~K.
Cycle stability ($R_h$) of materials was evaluated by capacity retaining rate after $n$ cycle: $R_h = (C_n/C_{max}) \times 100 \%$, where $C_n$ and $C_{max}$ are discharge capacities at the $n$-th cycle and maximum discharge capacity, respectively. 
\subsection{XPS Measurements \label{ssec:xps}}
The XPS spectra were measured at room temperature using the SPECS EA 10 PLUS energy spectrometer with Al-K$_\alpha$ radiation of 1486.6~eV. 
For the XPS valence band measurements we used a step size equal to 0.05~eV.
The energy spectra of the electrons were analyzed by a hemispherical analyzer
(FWHM$_{\mathrm{Mg-K_\alpha}}$ equals 0.8~eV for Ag 3$d_{5/2}$). 
Calibration of the spectra was performed according to Baer \textit{et al.}~\cite{baer_monochromatized_1975}.
The emission spectra of La, Mg and Ni were measured immediately after the sample transfer from the preparation chamber in a vacuum of 8~$\times$~10$^{-11}$~mbar. 
The 4$f_{7/2}$ peak of gold was situated at 84.0~eV and the Fermi level was located at binding energy 0~eV. 
The surface layer with impurities ($\sim10$~nm) of the studied bulk nanocrystalline samples could be removed using a SPECS ion gun etching system. 
We have used 3~keV Ar+ ion beam that was incident at an angle of 45\textdegree{} to the surface of the sample~\cite{smardz_xps_2008}.
The measurements were conducted following routine backing procedures ($T$~=~440~K) of the analysis chamber which made possible reaching a base vacuum of 4~$\times$~10$^{-11}$~mbar. 
Details of the XPS measurements can be found in Refs.~\cite{smardz_structure_2000, jurczyk_synthesis_2004, skoryna_xps_2015}.
\subsection{Density Functional Theory Calculations}
\label{sec:dft_methods}

%
%
The electronic band structure calculations of several phases related to (La,Mg)$_2$Ni$_7$ samples were carried out using the full-potential local-orbital minimum-basis scheme~\cite{koepernik_full-potential_1999,koepernik_self-consistent_1997}. 
The utilized FPLO5.00 version of the code is the latest public version allowing for the calculations within coherent potential approximation (CPA)~\cite{soven_coherent-potential_1967}.
The CPA was used to computationally model a chemical disorder as introduced to crystal structures of \lani{} by substituting Mg on La sites.
The form of CPA implementation in the FPLO5 prevents us from applying full-relativistic scheme and allows to use only a scalar-relativistic one.
Nevertheless, the constituent light elements (La, Mg, Ni) and the scope of this study justify the application of this approach.
Moreover, the additional full relativistic calculations made by us confirm this assumption.
For all calculations we use the local density approximation (PW92)~\cite{perdew_accurate_1992}, optimization of the basis, 16$^3$ \textbf{k}-mesh and convergence criterion of charge density 10$^{-6}$.
One exception is the \lamghalf{} in Ce$_2$Ni$_7$-type structure for which we had to limit the \textbf{k}-mesh down to 6$^3$.
%
Although the hexagonal \lani{} undergoes an antiferromagnetic transition at T$_N$~=~51~K~\cite{parker_magnetic_1983}, we consider only non-magnetic models as consistent with the non-magnetic states of the samples in room temperature.  
%
%
The theoretical photoemission spectra were obtained from the calculated densities of electronic states convoluted by Gaussian with a half-width ($\delta$) equal to 0.6~eV and scaled using the proper photoelectronic cross-sections for partial states~\cite{yeh_atomic_1985}.
This method we have successfully used before to calculate photoemission spectra of CeRh$_3$Si$_2$ and UGe$_2$ ~\cite{pikul_giant_2010,samsel-czekala_electronic_2011}.
The intention for convolution by Gaussian is to mimic a lifetime of hole states, an experimental broadening from instrumental resolution and thermal effects.
The value of $\delta = 0.6$~eV is selected as close to the value identified for the spectrometer used for measurements.
%

%
%
\begin{table*}[!ht]
\caption{\label{tab:dft_struct} 
The basic structural data of five compounds related to the considered (La,Mg)$_2$Ni$_7$ samples as used for \textit{ab initio} calculations.
}
\vspace{2mm}
\centering
\begin{tabular}{l|llllll}
\hline \hline 
phase        & prototype	& space group			& no.	  & $a$~(\AA{})	& $c$~(\AA{})	& Ref.\\
\hline
La$_2$Ni$_7$ & Ce$_2$Ni$_7$ 	& \textit{P}6$_3$/\textit{mmc}  & 194	  & 5.07	& 24.56		& this work\\
La$_2$Ni$_7$ & Gd$_2$Co$_7$	& \textit{R}$\bar{3}$\textit{m}	& 166	  & 5.056	& 36.98		& \cite{virkar_crystal_1969}\\
LaNi$_5$     & CaCu$_5$		& \textit{P}6/\textit{mmm}	& 191     & 5.010       & 3.972		& \cite{szajek_electronic_2000}\\
MgNi$_2$     & MgNi$_2$		& \textit{P}6$_3$/\textit{mmc} 	& 194	  & 4.8256      & 15.8323       & \cite{yartys_hydrogen-assisted_2015}\\ 
La$_2$O$_3$  & La$_2$O$_3$ 	& \textit{P}$\bar{3}$\textit{m}1& 164	  & 3.9381    	& 6.1361        & \cite{aldebert_etude_1979}\\  
\hline \hline
\end{tabular}
\end{table*}
%
%
%
\begin{figure}[ht]
\begin{center}
\includegraphics[trim = 0 10 0 10, clip, width=0.95\columnwidth]{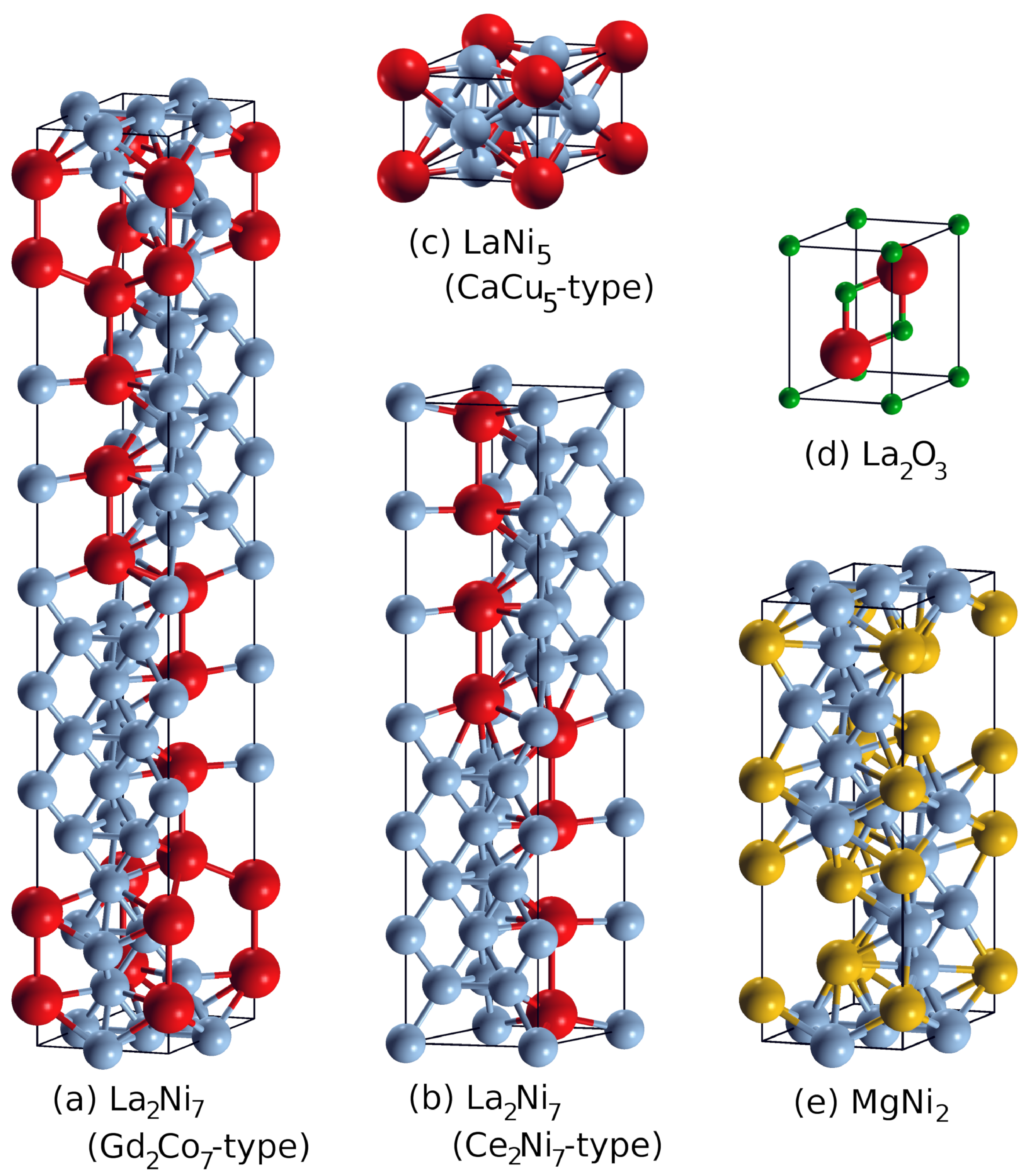} 
\end{center}
\caption{\label{fig:dft_struct}
The crystal structures of several compounds related to the considered (La,Mg)$_2$Ni$_7$ samples.
(a) La$_2$Ni$_7$ in Gd$_2$Co$_7$-type structure; (b) La$_2$Ni$_7$ in Ce$_2$Ni$_7$-type structure; (c) LaNi$_5$ in CaCu$_5$-type structure; (d) La$_2$O$_3$; (e) MgNi$_2$ (C36).
The different atoms are denoted by different colors as follows: La -- red, Ni -- blue, O -- green, Mg -- yellow.
}
\end{figure}

%
\begin{table}[!ht]
\caption{\label{tab:atomic_pos_ce2ni7} Crystallographic parameters of \lani{} in Ce$_2$Ni$_7$-type structure reproduced after Levin \textit{et al.}~\cite{levin_hydrogen_2004}. Lattice parameters $a$~=~5.0577, $c$~=~24.7336~\AA{}, space group \textit{P}6$_3$/\textit{mmc}.\\
}
\centering
\begin{tabular}{r|clll}
\hline \hline 
      atom& site &$x$ 		& $y$ 		& $z$ \\
\hline        
        La$_1$& 4$f$ &1/3	& 2/3		& 0.03013 \\
        La$_2$& 4$f$ &1/3	& 2/3		& 0.17422 \\
        Ni$_1$& 2$a$ &0		& 0		& 0 \\
        Ni$_2$& 4$e$ &0		& 0		& 0.16898 \\
        Ni$_3$& 4$f$ &1/3	& 2/3		& 0.83616 \\
        Ni$_4$& 6$h$ &0.8341	& 0.6682	& 1/4 \\
        Ni$_5$& 12$k$ &0.8350	& 0.6700	& 0.08806  \\ 
 \hline \hline        
\end{tabular}
\end{table}
\begin{table}[!h]
\caption{\label{tab:atomic_pos_gd2co7_hex} 
Crystallographic parameters of \lani{} in Gd$_2$Co$_7$-type structure -- space group \textit{R}$\bar{3}$\textit{m} (no.~166) hexagonal axes ($\alpha = \beta = 90^{\circ}, \gamma = 120^{\circ}$).
Lattice parameters $a$~=~5.056, $c$~=~36.98 \AA{} as measured by Virkar and Raman~\cite{virkar_crystal_1969}.
Atomic positions relaxed with use of the FPLO DFT code.\\
}
\centering
\begin{tabular}{r|clll}
\hline \hline 
      atom& site &$x$ 		& $y$ 		& $z$ \\
\hline        
        La$_1$& 6$c$ & 0        & 0        & 0.0501\\
        La$_2$& 6$c$ & 0        & 0        & 0.1459\\
        Ni$_1$& 3$b$ & 0        & 0        & 1/2\\
        Ni$_2$& 6$c$ & 0        & 0        & 0.2790\\
        Ni$_3$& 6$c$ & 0        & 0        & 0.3871\\
        Ni$_4$& 9$e$ & 1/2      & 0        & 0\\
        Ni$_5$& 18$h$ & 1/2     & -1/2     & 0.1077\\

 \hline \hline        
\end{tabular}
\end{table}
\begin{table}[!ht]
\caption{\label{tab:atomic_pos_gd2co7_rhomb} 
Crystallographic parameters of \lani{} in Gd$_2$Co$_7$-type structure -- space group \textit{R}$\bar{3}$/\textit{m} (no.~166) rhombohedral axes.
Lattice parameters $a$ = $b$ = $c$~~=~12.6676~\AA{} and $\alpha = \beta = \gamma = 23.02^{\circ}$ converted to rhombohedral representation from crystal refinements made by Virkar and Raman~\cite{virkar_crystal_1969}.
Atomic positions relaxed with use of the FPLO DFT code.\\
}
\centering
\begin{tabular}{r|clll}
\hline \hline 
      atom& site &$x$ 		& $y$ 		& $z$ \\
\hline        
        La$_1$& 6$c$ & 0.0501 & 0.0501 & 0.0501 \\
        La$_2$& 6$c$ & 0.1459 & 0.1459 & 0.1459 \\
        Ni$_1$& 3$b$ & -1/2 & -1/2 & -1/2\\
        Ni$_2$& 6$c$ & 0.2790 & 0.2790 & 0.2790 \\
        Ni$_3$& 6$c$ & 0.3871 & 0.3871 & 0.3871 \\
        Ni$_4$& 9$e$ & -1/2 & 1/2 & 0\\
        Ni$_5$& 18$h$ & -0.3931 & 0.1092 & -0.3931 \\     
 \hline \hline        
\end{tabular}
\end{table}
From first principles we calculate electronic structures of several phases identified by XRD in the considered (La,Mg)$_2$Ni$_7$ samples.
They are Ce$_2$Ni$_7$-type and Gd$_2$Co$_7$-type phases of La$_2$Ni$_7$, LaNi$_5$ and La$_2$O$_3$.
Furthermore we calculate \textit{ab initio} the bcc Ni and MgNi$_2$ (C36) phases.
Except for La$_2$Ni$_7$ with Gd$_2$Co$_7$-type structure, we use the lattice parameters and atomic coordinates as refined in experiments~\cite{virkar_crystal_1969, szajek_electronic_2000, yartys_hydrogen-assisted_2015,aldebert_etude_1979,levin_hydrogen_2004}.
In the face of lack in literature of the refined atomic positions for Gd$_2$Co$_7$-type structure of La$_2$Ni$_7$,
we have optimized for this phase the initial atomic coordinates of isostructural Y$_2$Ni$_7$ (Gd$_2$Co$_7$-type)~\cite{virkar_crystal_1969}. 
The optimized atomic positions with hexagonal and rhombohedral axes are collected in Tables~\ref{tab:atomic_pos_gd2co7_hex} and  \ref{tab:atomic_pos_gd2co7_rhomb}.
They are in good agreement with the  atomic positions refined with Rietveld analysis for isostructural \lamghalf{} (Gd$_2$Co$_7$-type)~\cite{zhang_mgni_2007}.
Furthermore, Tab.~\ref{tab:atomic_pos_ce2ni7} presents the crystallographic parameters of \lani{} in Ce$_2$Ni$_7$-type structure as measured by Levin \textit{et al.}~\cite{levin_hydrogen_2004} and used by us for calculations.
The collection of basic structural data for other phases taken into account is presented in Tab.~\ref{tab:dft_struct}.
The considered crystal structures are also shown in Fig.~\ref{fig:dft_struct}.
For \lamghalf{} systems modeled in CPA we used the experimental lattice parameters.
As the FPLO code precludes optimization if the CPA applied, the atomic positions were used as for corresponding \lani{} structures.
For visualization of crystal structures the XCrySDen computer code~\cite{kokalj_computer_2003} was used.
\section{Results and Discussion}
%
%
The \lamgni{} alloys powders ($x$~=~0.00, 0.25, 0.50 and 0.75) were prepared by the MA and annealing process.
In this section we will present the results and discussion of structural and valence band characterization, electrochemical measurements and \textit{ab initio} calculations.
\subsection{Structural Characterization}
\label{subsec:struct_charact}
\begin{figure}[ht]
\begin{center}
\includegraphics[trim = 200 220 230 210,clip,width=0.75\columnwidth]{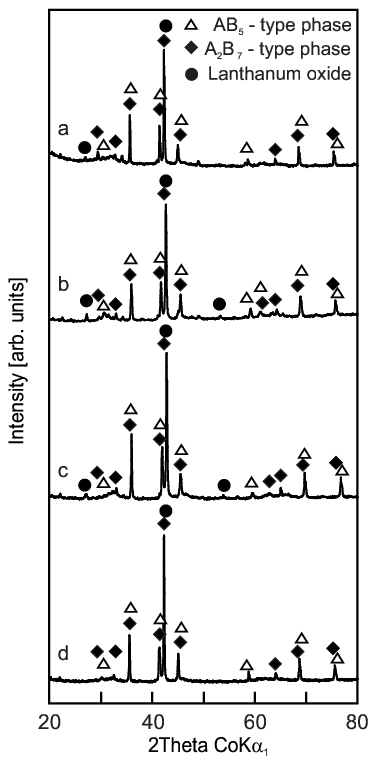} 
\end{center}
\caption{\label{fig:xrd}
The XRD spectra of \lamgni{} after 48~h mechanical alloying and annealing at 1123~K for 0.5~h. 
(a) $x$~=~0.00, (b) $x$~=~0.25, (c) $x$~=~0.50 and (d) $x$~=~0.75.
}
\end{figure}
\begin{table}[!ht]
\caption{
\label{tab:lattice_and_electro} 
XRD resolved lattice parameters of Ce$_2$Ni$_7$-type phase and electrochemical properties of \lamgni{} alloys. 
Maximum discharge capacities ($C_{max}$), discharge capacities at 30 cycles ($C_{30}$) and cycle stability ($R_h$).\\
}
\centering
\begin{tabular}{l|ccc|ccc}
\hline \hline
$x$	& $a$ 	& $c$	& $V$		& $C_{max}$	& $C_{30}$& $R_h$\\
	&\AA{}	&\AA{}	&\AA{}$^3$	& mAh/g		& mAh/g	&	\\
\hline
0.00	& 5.07	& 24.56	& 546.58	& 76		& 30	& 39 \\
0.25	& 5.06	& 24.49	& 543.56	& 213		& 146	& 69 \\
0.50	& 5.04	& 24.20	& 533.21	& 304		& 204	& 67 \\
0.75	& 5.03	& 24.22	& 531.26	& 237		& 226	& 95 \\
\hline \hline
\end{tabular}              
\end{table} 
After 48~h of MA the originally sharp diffraction peaks (not shown) of the constituent elements La, Mg and Ni lost their intensity.
The subsequent heat treatment at 1123~K for 0.5~h (under high purity argon atmosphere) induces crystallization in the \lamgni{} samples.
Accordingly to XRD patterns, in the final \lamgni{} samples no single-element phases of La, Mg nor Ni was present.
The XRD spectra of the resultant \lamgni{} alloys are presented in Fig.~\ref{fig:xrd}.
Table~\ref{tab:lattice_and_electro} presents how the lattice parameters of Ce$_2$Ni$_7$-type phase decrease with Mg concentration.
%
%
\begin{table}[!ht]
\caption{\label{tab:polyphase_composition} 
The composition of \lani{} and \lamghalf{} samples (in wt.\% of the constituent phases).\\
}
\centering
\begin{tabular}{lcc}
\hline \hline 
phase				&\lani{}& \lamghalf{}	\\
\hline
La$_2$Ni$_7$ (Ce$_2$Ni$_7$-type)& 46.1	& 16.0		\\
La$_2$Ni$_7$ (Gd$_2$Co$_7$-type)& 46.1	& 72.9		\\
LaNi$_5$     			& 5.6	& 8.0		\\
La$_2$O$_3$   			& 2.2	& 3.1		\\
\hline \hline        
\end{tabular}
\end{table}
Based on the peaks intensities analysis we conclude that the resultant \lani{} sample has the multi-phase character and 
consists of Ce$_2$Ni$_7$-type (46.1~wt.\%) and Gd$_2$Co$_7$-type (46.1~wt.\%) structures of La$_2$Ni$_7$ phase, LaNi$_5$ (5.6~wt.\%) and La$_2$O$_3$ (2.2~wt.\%) phases.
In the \lamghalf{} sample ($x$~=~0.5) the Gd$_2$Co$_7$-type of La$_2$Ni$_7$ phase (72.9~wt.\%) prevails with a smaller share of Ce$_2$Ni$_7$-type (16.0~wt.\%)  of La$_2$Ni$_7$ and additions of LaNi$_5$ (8.0~wt.\%) and La$_2$O$_3$ (3.1~wt.\%).
For \lamghalf{} sample Zhang \textit{et al.} reported composition of about 60\% Gd$_2$Co$_7$-type and 40\% Ce$_2$Ni$_7$-type phase after annealing at above 1000~K, which stays in relatively good agreement with our result.~\cite{zhang_mgni_2007}  
The data on phase composition of \lani{} and \lamghalf{} samples are gathered in Tab.~\ref{tab:polyphase_composition}.
The collection of basic structural parameters of the \lamgni{} constituent phases can be found in Tab.~\ref{tab:dft_struct} and  Fig.~\ref{fig:dft_struct}
gives their graphical representations.
%
%
The Ce$_2$Ni$_7$-type and Gd$_2$Co$_7$-type structures of La$_2$Ni$_7$ phase are closely related.
They crystallize in \textit{P}6$_3$/\textit{mmc} and \textit{R}$\bar{3}$\textit{m} space groups, respectively. 
The crystal structures are highly anisotropic and can be understand as stacking of LaNi$_5$ and LaNi$_2$ unit blocks along the $c$ axis~\cite{crivello_structural_2011}.
An almost equal concentration of the Ce$_2$Ni$_7$-type and Gd$_2$Co$_7$-type structures in our \lani{} sample can be understand based on by nearly identical heat of formation of these structures~\cite{crivello_structural_2011}.
The heat of formation analysis indicates also that the in case of \lamghalf{} sample a preferred atomic positions of Mg/La substitutions are 4$f_1$ in case of Ce$_2$Ni$_7$-type and 6$c_2$ in Gd$_2$Co$_7$-type structure~\cite{crivello_structural_2011}.
The complete tables with atomic coordinates of \lani{} phases can be found in Sec.~\ref{sec:dft_methods}.
The detailed crystallographic structures of metal hydrides La$_2$Ni$_7$D$_x$~\cite{yartys_novel_2006} and La$_{1.5}$Mg$_{0.5}$Ni$_7$D$_x$~\cite{denys_mg_2008} (of Ce$_2$Ni$_7$-type structure) have been refined.
\subsection{Electrochemical Measurements}
\begin{figure}[ht]
\begin{center}
\includegraphics[trim = 5 165 100 130,clip,width=0.95\columnwidth]{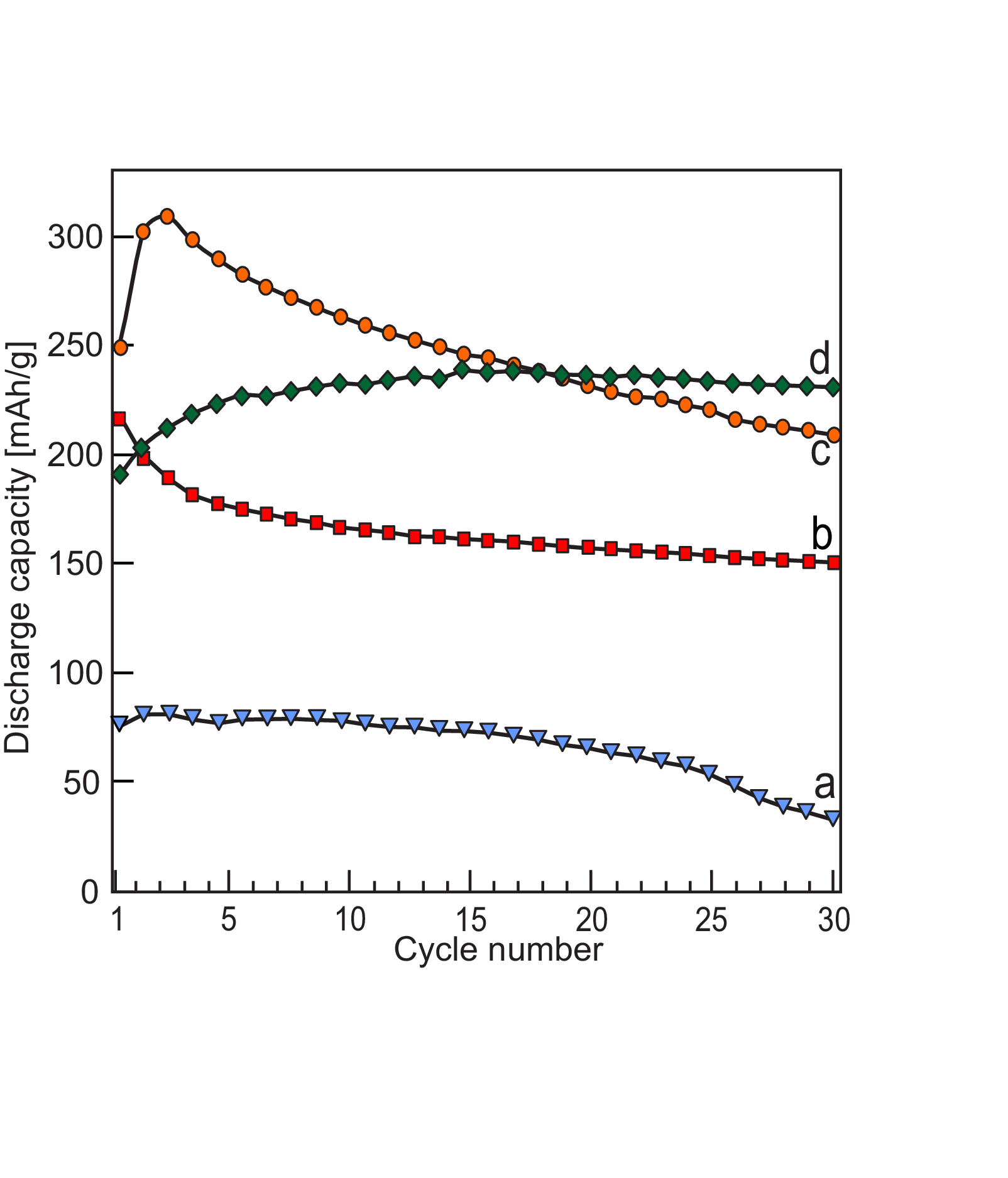} 
\end{center}
\caption{\label{fig:discharge}
Discharge capacities as a function of cycle number of \lamgni{}. (a) $x$~=~0.00, (b) $x$~=~0.25, (c) $x$~=~0.50 and (d) $x$~=~0.75.
}
\end{figure}
Results of electrochemical measurements for \lamgni{} alloys ($x$~=~0.00, 0.25, 0.50 and 0.75) are presented in Fig.~\ref{fig:discharge} as
discharge capacities in function of cycle number.
Furthermore, the measured electrochemical characteristics of the materials are summarized in Tab.~\ref{tab:lattice_and_electro}. 
It is concluded that the best activation properties has the unmodified \lani{} alloy and
an addition of Mg causes only deterioration of the activation properties. 
The activation capability of an alloy electrode is related to the change of the internal energy of the hydride system before and after hydrogen absorption~\cite{zhang_electrochemical_2015}.
The electrochemical properties of the alloy depends on multiple factors, involving its crystal structure, phase composition and microstructure. 
Most of the electrodes display maximum discharge capacity ($C_{max}$) during the first three cycles. 
$C_{max}$ is increasing with Mg content to reach the highest value (304~mAh/g) for $x$~=~0.5 (\lamghalf). 
Further increase of Mg concentration causes decrease of $C_{max}$. 
The same tendency was observed in pressure-composition-temperature (PCT) tests~\cite{balcerzak_hydrogenation_2017}.
The degradation of discharge capacity in La-Mg-Ni alloys originates from forming and increasing of Mg(OH)$_2$ and La(OH)$_3$ surface layers. 
These oxides hinder the hydrogen atoms from diffusion in or out in alkaline solutions~\cite{dornheim_hydrogen_2007}.
\begin{figure}[ht]
\centering
\includegraphics[trim = 50 70 85 80,clip,width=\columnwidth]{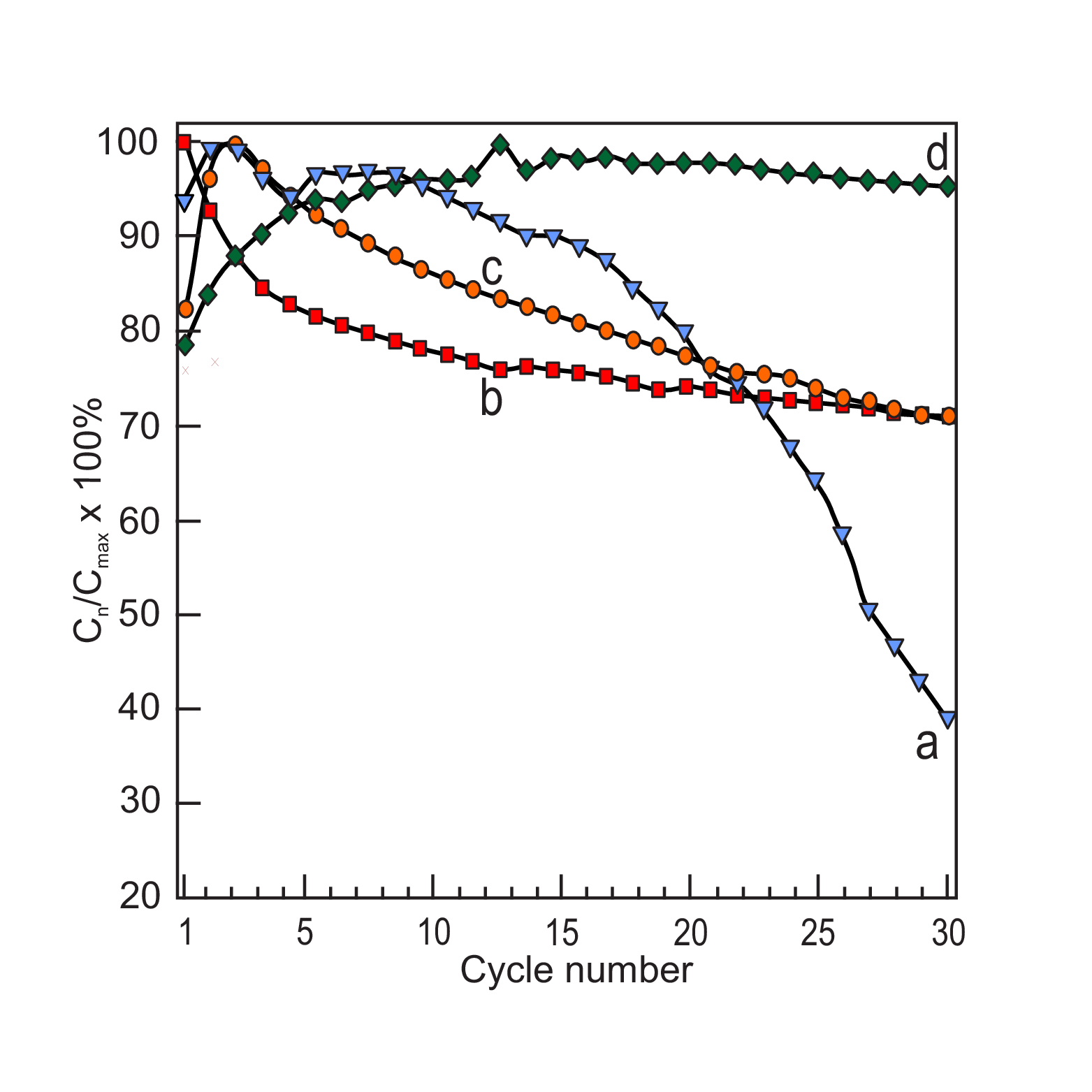}
\caption{\label{fig:cyclic}
Cyclic stability of electrodes prepared from \lamgni{} alloys. 
(a) $x$~=~0.00, (b) $x$~=~0.25, (c) $x$~=~0.50 and (d) $x$~=~0.75.
}
\end{figure}
Partial substitution of La by Mg causes increase of cycle stability of electrodes, see Fig.~\ref{fig:cyclic}. 
The best cycle stability was obtained for the material where $x$~=~0.75. 
It is due to the phase composition of this material. 
In La$_{1.25}$Mg$_{0.75}$Ni$_7$ alloy, except the main (La,Mg)$_2$Ni$_7$ phase, the LaNi$_5$ phase is observed,
which possesses much higher electrochemical cycle stability than (La,Mg)$_2$Ni$_7$ phase~\cite{yanghuan_electrochemical_2013}.
Taking into account the $C_{max}$, a sample with the most promising composition is \lamghalf{} ($x$~=~0.5) and further investigations presented in this work will focus on it. 

\subsection{XPS Measurements}
It has been shown that the formation of metal hydrides follows the rules of semi-empirical models~\cite{bouten_heats_1980, griessen_heat_1988},
which characteristic parameters depend on the features of the valence band.
One of the experimental technique of characterization of the valence band is the X-ray photoelectron spectroscopy (XPS).
The analysis of the valence band provides information on electronic structure together with features of crystal structure and microstructure.
These electronic and geometric factors affect the metal-hydrogen interaction.
\begin{figure}[ht]
\centering
\includegraphics[trim = 280 100 10 90,clip,height=1.0\columnwidth,angle=270]{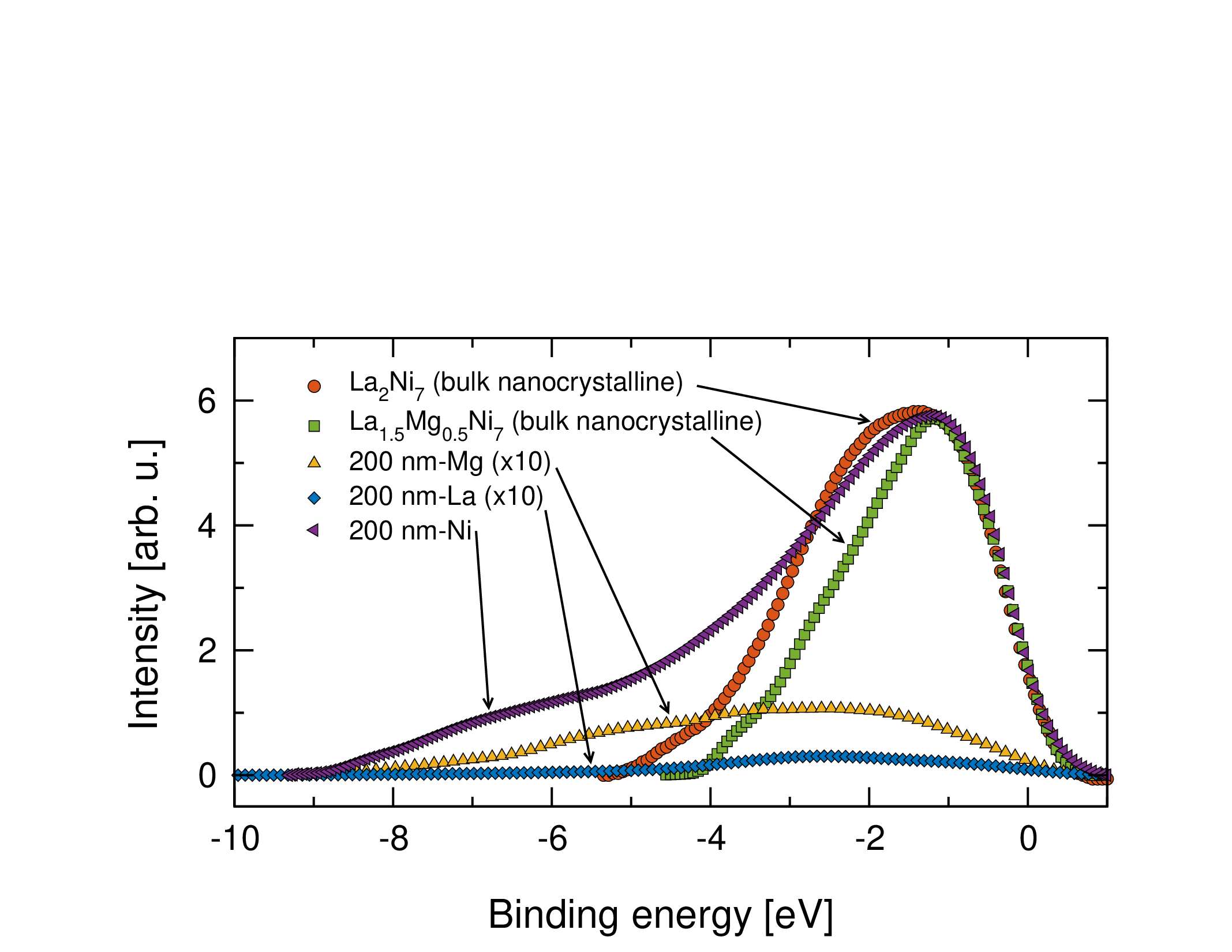}
\caption{\label{fig:xps}
The XPS valence band spectra (Al-K$_\alpha$) of bulk nanocrystalline \lani{} and  \lamghalf{} alloys, together with valence band spectra of \textit{in situ} prepared pure La, Mg and Ni thin films presented for comparison.
}
\end{figure}
The measured XPS valence band spectra are presented in Fig.~\ref{fig:xps}.
Instead of investigating the whole range of \lamgni{} compositions with the XPS, 
we focus on the \lamghalf{} sample which, among the investigated series, exhibited the highest value of discharge capacity ($C_{max}$~=~304~mAh/g).
As reference we study with XPS also the bulk nanocrystalline \lani{} sample and \textit{in situ} prepared La, Mg and Ni thin films.
%
%
The XPS spectra of the elemental thin films suggest that the main contribution to the \lani{} and \lamghalf{} valence bands comes from the 3$d$ electrons of Ni.
The full width at half maximum (FWHM) of the \lani{} and \lamghalf{} valence bands are about 2.7~eV and 2.2~eV, respectively.
The narrowing of the \lamghalf{} valence band could be explained by increase of amount of the La oxide phase in the surface layer.
This concept will be investigated in the next section covering the \textit{ab initio} results. 
Note, that the XPS signal originates mainly from the surface layer (3--5~nm)~\cite{smardz_xps_2008,smardz_electronic_2008}.
Furthermore, in Ref.~\cite{smardz_xps_2008} we showed that La atoms segregate to the surface of the nanocrystalline LaNi$_5$-type alloys and form stable oxides under ambient conditions. 
The oxidized top layer is depth-limited and forms a good protective layer against further oxidation under ambient conditions. 
As described in Sec.~\ref{ssec:xps}, we have removed only of about 10~nm surface impurity layer which mainly consisted of carbon and oxygen. 
However, the subsurface layer composed of a significant amount of La oxides was not removed. 
In the case of nanocrystalline \lani{} alloy the content of the oxide in the subsurface layer is significantly lower probably due to negligible La segregation effect. 
The above result is in good agreement with average La$_2$O$_3$ content estimated from the XRD experiment, see Tab.~\ref{tab:polyphase_composition}. 
Therefore, the valence band of the nanocrystalline \lamghalf{} is narrower compared to that measured for the \lani{} alloy. 
Theoretical XPS valence bands calculated for perfect crystalline materials without segregation and oxidation effects are practically the same, see Sec.~\ref{ssec:valence_band}. 
Furthermore, looking from other perspective, such an additional surface oxide layer could influence hydrogen absorption and electrochemical properties of the studied material, as has been recently reported for the well defined Mg/Pd bilayers with La interlayer~\cite{liu_superior_2014}.
\subsection{Density Functional Theory Calculations}
%
%
%
In DFT calculations, similar like in the XPS measurements, we focus on the \lamghalf{} sample with the highest $C_{max}$ instead of considering the whole concentration range of Mg.
It has been shown that the inhomogeneous \lamghalf{} sample consists, among others, of \lamghalf{} in Ce$_2$Ni$_7$-type and Gd$_2$Co$_7$-type phases, LaNi$_5$ and La$_2$O$_3$, see Tab.~\ref{tab:polyphase_composition}.
 Furthermore, for the neighboring composition (LaMgNi$_7$)  Crivello \textit{et al.}~\cite{crivello_structural_2011} suggested an energetically stable state with LaNi$_5$ phase coexisting with MgNi$_2$.
The above observations are the reason why we decide to investigate theoretically the several phases mentioned above.
That list has been extended by the reference phases: Ce$_2$Ni$_7$-type and Gd$_2$Co$_7$-type of \lani{} and fcc Ni.
This analysis is designed to answer the questions of the energetic stability of the particular phases and to better understand the electronic properties and the nature of \lamghalf{} valence band (studied in this work also by XPS).
The basic structural data of the considered phases are collected in Tabs.~\ref{tab:dft_struct}, \ref{tab:atomic_pos_ce2ni7}, \ref{tab:atomic_pos_gd2co7_hex} and \ref{tab:atomic_pos_gd2co7_rhomb}.
For graphical representation of crystallographic structures see Fig.~\ref{fig:dft_struct}.
\subsubsection{Site preference and phase stability}
%
%
Our calculations on LaNi$_5$, La$_2$O$_3$, MgNi$_2$ and fcc Ni are dedicated to resolve the measured XPS valence band of inhomogeneous samples.
However nearly 90\% of the \lamghalf{} composition consists of Ce$_2$Ni$_7$-type and Gd$_2$Co$_7$-type phases, see Tab.~\ref{tab:polyphase_composition}.
This is why the latter phases are the main subject of our DFT study.
Although the basic \lani{} phases can be simulated based on the simple ordered structures, 
the modeling of the systems with chemical disorder is more demanding.
Previously the \lamgni systems with Mg have been modeled based on the ordered compound method~\cite{crivello_first_2015,crivello_structural_2011}.
In this work the chemical disorder introduced by Mg substitution on La sites is considered based on the coherent potential approximation (CPA).
One of the advantages of CPA method is that it allows for any arbitrary concentration and not only $x$~=~0.5, 1.0 or 1.5.
We have applied the CPA before to model Nb and Ti substitutions in YCo$_2$~\cite{sniadecki_induced_2014}.

%
%
Experimental observations of phase composition indicate about 46 wt.\% shares of hexagonal Ce$_2$Ni$_7$-type and rhombohedral Gd$_2$Co$_7$-type phases in our  \lani{} samples, see Tab.~\ref{tab:polyphase_composition}.
The formation of significant amounts of both phases suggests small energy difference between them,
which finds a confirmation in nearly identical heats of formation calculated by Crivello \textit{et al.}~\cite{crivello_structural_2011} for \lani{} in two above structure types.
So good energy balance between both types of structures comes form a close relationship between them, consisting both of LaNi$_2$ and LaNi$_5$ unit blocks~\cite{crivello_structural_2011}. 
The total energy comparison made by us indicates that both structure types have nearly the same total energies with a small preference of about 10~meV/atom in favor of Gd$_2$Co$_7$-type which is in good agreement with the previous results~\cite{crivello_structural_2011}.
Furthermore, the total energy comparison between hexagonal and rhombohedral phases of La$_5$Ni$_{19}$~\cite{liu_enhanced_2015} indicates a very small preference of about 1.4~meV/atom in favor of hexagonal phase.
The above results of small energy differences together with the heats of formation calculated by Crivello \textit{et al.}~\cite{crivello_structural_2011} suggest that in the whole La-Ni-Mg system might be difficult to find preference of hexagonal or rhombohedral-type structures.

%
In \lamghalf{} samples with 25\% doping of Mg on La sites also occur hexagonal Ce$_2$Ni$_7$-type and rhombohedral Gd$_2$Co$_7$-type phases.
In order to describe site preferences for Mg, both structure types have to be considered.
%
%
Zhang \textit{et al.}~\cite{zhang_structure_2007} determined from Rietveld analysis that the Mg atoms are located only on the 4$f$ La$_1$ sites (in Laves unit blocks LaNi$_2$) of hexagonal \lamghalf{} phase (Ce$_2$Ni$_7$-type), see Tab.~\ref{tab:atomic_pos_ce2ni7}. 
The same conclusions were drawn once again by Denys \textit{et al.}~\cite{denys_mg_2008}.
The beneficial conditions for Mg substitution on La 4$f_1$ site ($\Delta E \sim 60$~meV/atom) were also predicted with DFT~\cite{crivello_structural_2011}.
We have once again attempted to determine the Mg site preferences in \lamghalf{} (Ce$_2$Ni$_7$-type), but this time based on the CPA models.
We consider Mg atoms located at 4$f_1$ sites, 4$f_2$ sites or randomly distributed between them. 
Our total energy calculations show that the most stable configuration is when the Mg atoms are located at La 4$f_1$ sites, and the least probable is random distribution of dopants between 4$f_1$ and 4$f_2$ sites. 
The difference in total energy between these configurations is about 70 meV/atom, which is relatively large value and suggests rather strong preference, even above room temperature.

%
For Gd$_2$Co$_7$-type phase of \lamghalf{} Zhang \textit{et al.}~\cite{zhang_mgni_2007} determined experimentally that Mg atoms prefer to occupy the La$_2$ 6$c_2$ sites, however together with minor tendency towards the La$_1$ 6$c_1$ sites, see Tabs.~\ref{tab:atomic_pos_gd2co7_hex} and \ref{tab:atomic_pos_gd2co7_rhomb}.
Based on the DFT calculations Crivello \textit{et al.}~\cite{crivello_structural_2011} also indicated for Gd$_2$Co$_7$-type phase of \lamghalf{} the 6$c_2$ sites preference of Mg ($\Delta E \sim 60$~meV/atom).

%
In case of \lamghalf{} sample our diffraction refinement indicates 16\% of Ce$_2$Ni$_7$-type and  72.9\% of Gd$_2$Co$_7$-type phase, see Tab.~\ref{tab:polyphase_composition}.
Previously Zhang \textit{et al.} reported also the composition of about 60\% Gd$_2$Co$_7$-type and 40\% Ce$_2$Ni$_7$-type phase in \lamghalf{} sample~\cite{zhang_mgni_2007}.
Similar like for \lani{} samples, the formation of both phases suggests small energy difference between them.
Based on DFT calculations Crivello \textit{et al.}~\cite{crivello_structural_2011} found nearly identical heats of formation for both  hexagonal and rhombohedral ground state structures of \lamghalf{}.
Our total energy calculations comparing \lamghalf{} phase of Ce$_2$Ni$_7$-type and Mg at 4$f_1$ sites with \lamghalf{} of Gd$_2$Co$_7$-type phase and Mg at 6$c_2$ sites gives energy difference of about 100 meV/atom with a preference of Gd$_2$Co$_7$-type phase.

Concluding, the \lani{} and \lamghalf{} phases are stable in both hexagonal Ce$_2$Ni$_7$-type and rhombohedral Gd$_2$Co$_7$-type structures.
In \lamghalf{} phase the Mg site preferences are  4$f_1$ in hexagonal and  6$c_2$ in rhombohedral structure types.
These two ground state \lamghalf{} structures, together with two \lani{} structures will be considered in the next section devoted to the analysis of the valence band.

\subsubsection{Valence band analysis \label{ssec:valence_band}}
\begin{figure}[ht]
\centering
\includegraphics[trim = 110 80 30 20,clip,width=\columnwidth]{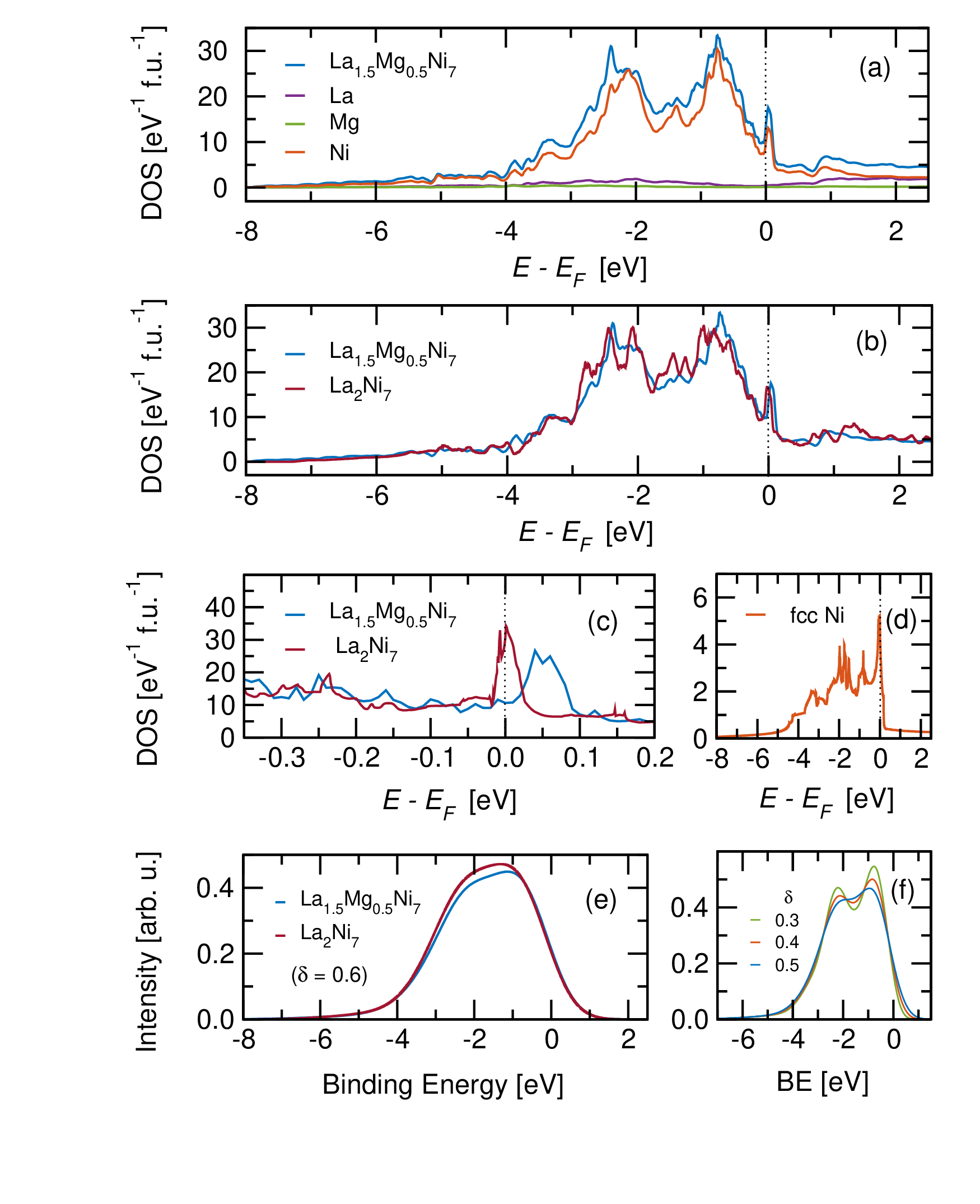}
\caption{\label{fig:dos_xps}
(a) Density of states (DOS) for \lamghalf{} (hexagonal Ce$_2$Ni$_7$-type) together with element-resolved contributions,
(b),(c) DOS for \lani{} and \lamghalf{} (both Ce$_2$Ni$_7$-type), 
(d) DOS for fcc Ni,
(e) calculated XPS spectra of \lani{} and \lamghalf{} (both Ce$_2$Ni$_7$-type),
(f) calculated XPS spectra of \lamghalf{} Ce$_2$Ni$_7$-type for different $\delta$ parameters.
}
\end{figure}

Our valence band DFT calculations are dedicated both as (1) interpretation for experimental XPS and (2) self-standing results shedding light on the electronic structure of the \lamghalf{} and at the same time filling the gap in understanding of the system.   
%
%
%
\begin{figure}[h]
\begin{center}
\includegraphics[trim = 65 20 40 120,clip,width=\columnwidth]{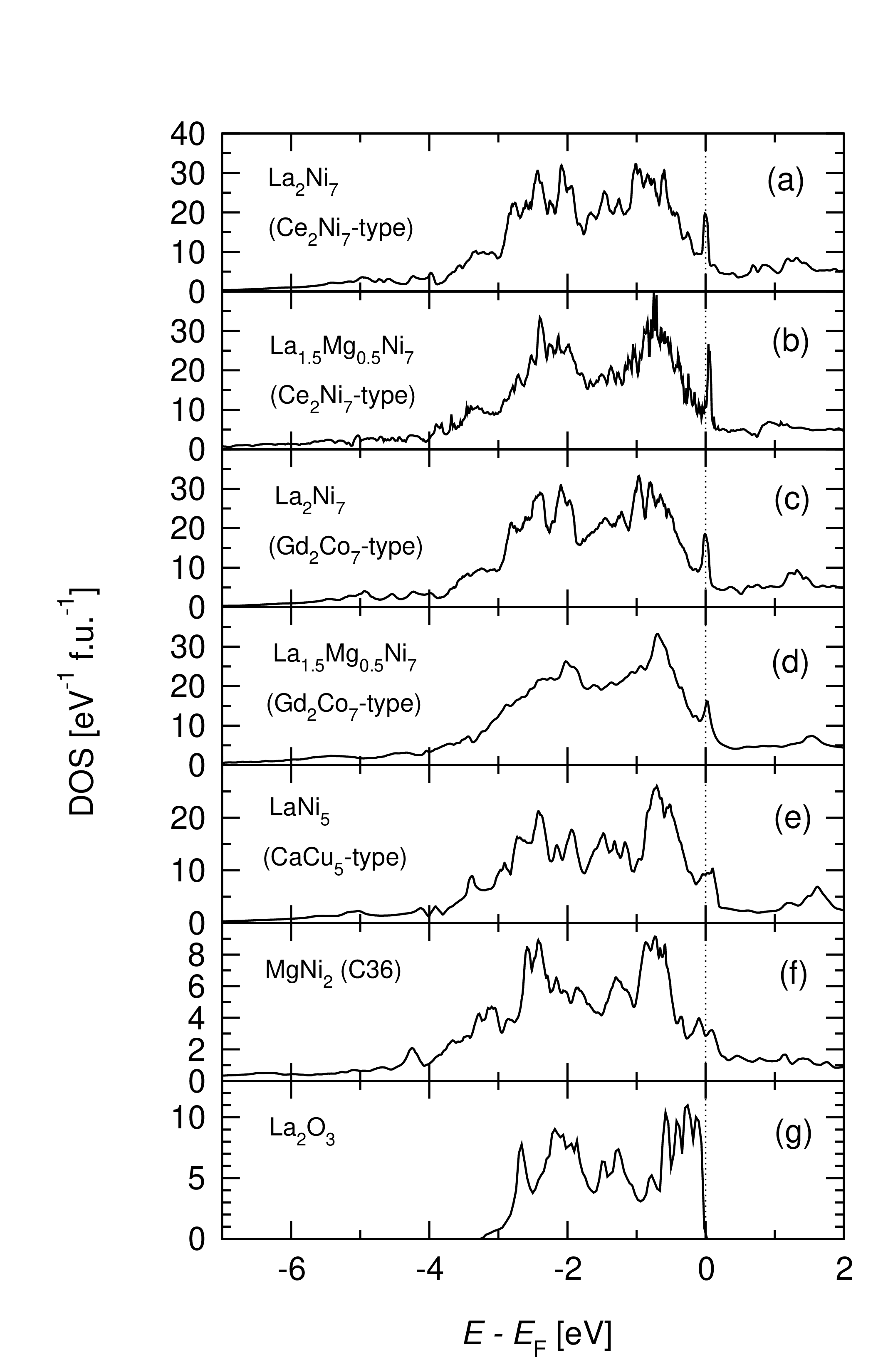} 
\end{center}
\caption{\label{fig:dos_comp}
Densities of states (DOS) calculated for (a) \lamgni{} (Ce$_2$Ni$_7$-type), (b) \lamghalf{} (Ce$_2$Ni$_7$-type), (c) \lani{} (Gd$_2$Co$_7$-type), (d) \lamghalf{} (Gd$_2$Co$_7$-type), (e) LaNi$_5$, (f) MgNi$_2$ (C36) and (g) La$_2$O$_3$ phases.  
}
\end{figure}
The valence bands of both \lani{} and \lamghalf{} phases in hexagonal Ce$_2$Ni$_7$-type and rhombohedral Gd$_2$Co$_7$-type structures are presented in Fig.~\ref{fig:dos_comp}. 
There is not much differences between the densities of states (DOS) calculated for hexagonal and rhombohedral phases.
The main part of the valence bands start in each case about -5.0~eV and finish soon after Fermi energy (\textit{E}$_\mathrm{F}$).
In every case two main maxima (at -2.3 and -0.8 eV) and characteristic sharp peak around \textit{E}$_\mathrm{F}$ are observed.
The main contributions to the valence bands come from the Ni 3$d$ bands, see Fig.~\ref{fig:dos_xps}(a), and resemble the valence band of the bcc Ni, see Fig.~\ref{fig:dos_xps}(d).

%
As Mg atoms have one less valence electron than La ones 
the substitution of Mg in place of La in \lani{} slightly depopulates the valence band.
In terms of rigid band approximation the Fermi level (\textit{E}$_\mathrm{F}$) of \lamghalf{} shifts a little bit left towards the higher binding energies, see Figs.~\ref{fig:dos_xps}(b) and (c).
It is observed that the Ni atoms in \lamghalf{} have about 0.4 less of 3$d$ electrons per band than in \lani{}. 
In case of \lani{} even a small change in valence band occupation may significantly affect the conduction properties of the material, 
as the sharp peak is located at the Fermi level.
In fact, the Mg substitution in \lamghalf{} reduces the value of DOS(\textit{E}$_\mathrm{F}$) from about 30 to 11~states/(eV\,f.u.) for hexagonal Ce$_2$Ni$_7$-type structure and from about 29 to 14~states/(eV\,f.u.) for the rhombohedral Gd$_2$Co$_7$-type structure. 
These values of DOS(\textit{E}$_\mathrm{F}$) correspond to the parameter $\gamma$ in linear term of specific heat of 72 and 25~mJ/(mol\,K$^2$) for hexagonal and 68 and 34~mJ/(mol\,K$^2$) for rhombohedral structure, respectively.
The above values of DOS(\textit{E}$_\mathrm{F}$) and $\gamma$ consist mainly of contributions from Ni electrons (above 95\%).

%
%
%
In Fig.~\ref{fig:xps} we present the measured XPS valence band spectra of bulk nanocrystalline \lani{} and \lamghalf{} alloys.
As the XPS resolution is limited the observed spectra are rather fuzzy.
The FWHM of the \lani{} and \lamghalf{} valence bands are about 2.7~eV and 2.2~eV, respectively.
In Fig.~\ref{fig:dos_xps}(e) and (f) we show the calculated XPS spectra of \lani{} and \lamghalf{} based on the DOS of hexagonal Ce$_2$Ni$_7$-type structures.
The calculated valence band width FWHM of \lani{} of about 2.8~eV stays in very good agreement with the measurements.
As the addition of Mg shifts the DOS only slightly, the resultant theoretical XPS spectra of \lamghalf{} are then also nearly identical as for \lani{} and its FWHM is about 2.7~eV.
This number is much bigger than experimental 2.2~eV for \lamghalf{} sample and from theory we can conclude that the narrowing of experimental XPS of \lamghalf{} in respect to \lani{} cannot be explained based on the single phase arguments.
Trying to find the origins for such narrowing we considered from DFT the valence bands of LaNi$_5$, La$_2$O$_3$, bcc Ni and MgNi$_2$ (C36) phases, which all may be part of the \lamghalf{} sample, see Figs.~\ref{fig:dos_comp}(e-g) and \ref{fig:dos_xps}(d).
The comparison of the calculated DOS's indicates that all considered structures with Ni have very similar width of the valence band.
Therefore the experimentally observed narrowing of the \lamghalf{} valence band (FWHM~=~2.2~eV) cannot be explained based on the presence of other phases containing Ni.
However, the effect of narrowing can be related to the presence of the oxide phase La$_2$O$_3$, which valence band showed in Fig.~\ref{fig:dos_comp}(g) is noticeably narrower from the remaining Ni compounds.
In fact, the presence of 3.1~wt.\% of La$_2$O$_3$ phase have been observed in \lamghalf{} sample.
The theoretical XPS spectra presented in Fig.~\ref{fig:dos_xps}(e) were obtained from the partial DOS's convoluted by Gaussians with a half-width $\delta$ equal to 0.6~eV.
For smaller $\delta$ the XPS spectra would reveal more features of valence band, see Fig.~\ref{fig:dos_xps}(f).
The broadening parameter $\delta$ of our XPS equipment is however only about 0.4--0.5~eV. 
Therefore, the double-peak structure observed in theoretical XPS for $\delta = 0.3$~eV, see Fig.~\ref{fig:dos_xps}(f), could not be directly observed experimentally, see Fig.~\ref{fig:xps}. 
The theoretical XPS spectra are also calculated for the perfect \textit{single-crystalline} materials, 
when the experimental XPS spectra of the real nanocrystalline samples are further obscured due to broadening coming from a complex microstructure of the material.
The broadening of the experimental valence band of the nanocrystalline La$_2$Ni$_7$  alloy could be explained by a strong deformation of the nanocrystals in mechanically alloyed and heat treated samples~\cite{smardz_xps_2012}.
For such nanocrystalline samples the interior of the nanocrystal is constrained and the distances between atoms located at the grain boundaries expanded. 
In Ref.~\cite{skoryna_xps_2015} we have compared the XPS valence bands of mechanically alloyed bulk nanocrystalline LaNi$_{4.2}$Al$_{0.8}$ and high-purity poly- and nanocrystalline LaNi$_4$Al thin films. 
For this case we have also observed the valence band broadening in the thin film and bulk nanocrystalline material in relation to the polycrystalline sample.

\section{Summary and Conclusions}

%
We studied the effect of substitution La by Mg on electrochemical and electronic properties in \lamgni{} phases.
The experimental preparation and characterization part was followed by the detailed DFT study.
%
%
First, the series of \lamgni{} samples (with $x$~=~0.00, 0.25, 0.50 and 0.75) was obtained by mechanical alloying.
Next, they were characterized by XRD, XPS and electrochemical measurements.
We concluded that the resultant \lani{} samples have the multi-phase character and 
consist mainly of Ce$_2$Ni$_7$-type and Gd$_2$Co$_7$-type structures of (La,Mg)$_2$Ni$_7$, with minor contributions of LaNi$_5$ and La$_2$O$_3$ phases.
Best electrochemical properties (maximum discharge capacity) were identified for \lamghalf{} sample.
This composition was then a subject of the XPS investigations covering the valence band and reference samples of La, Mg, Ni and \lani{}.
The main contribution from Ni electrons to valence band was established.
The observed narrowing of the \lamghalf{} valence band in respect to \lani{} is supposed to be caused by increase of the lanthanum oxide phase concentration.
%
%
With DFT calculations we focused also on the \lamghalf{} sample, 
although we considered also the other phases identified by XRD in our multi-phase samples.
Our \textit{ab initio} analysis was intended to investigate the energetic stability of the \lamgni{} phases and to provide information on the \lamghalf{} valence band.
In face of lack of experimental data, first we determined the atomic positions for Gd$_2$Co$_7$-type \lani{}-phase by structure optimization.
We identified the Gd$_2$Co$_7$-type phases as only slightly more stable than the Ce$_2$Ni$_7$-type for both \lani{} and \lamghalf{} phases.  
For modeling of the Mg substitution in \lamghalf{} phases we used the coherent potential approximation. 
We conclude from calculations that the Mg atoms prefer to occupy La 4$f_1$ sites for Ce$_2$Ni$_7$-type \lamghalf{} phase and the La 6$c_2$ positions for Gd$_2$Co$_7$-type phase.
These results stay in good agreement with the previous experimental and theoretical statements.
The stability analysis was followed by the valence band investigations -- primarily dedicated to interpret the experimental XPS spectra.
It confirmed that the \lamghalf{} valence band consist mainly of the contribution from Ni electrons.
The Mg substitution in place of La in \lamgni{} only slightly depopulates the valence band. 
However, as the strongly localized states were identified near the Fermi energy level, even small shift of Fermi level may affect the conductivity of the material.

\section*{Acknowledgements}
Work supported by the National Science Centre Poland under the decision DEC-2014/15/B/ST8/00088. 
Part of the computations was performed on the resources provided by the Poznań Supercomputing and Networking Center (PSNC).

\end{sloppypar}

\bibliography{la2ni7}

\end{document}